\documentclass[prb,twocolumn,showpacs,preprintnumbers]{revtex4}
\usepackage{graphics}
\usepackage{graphicx}
\usepackage{dcolumn}
\usepackage{bm}
\usepackage{epsfig}
\begin{document}
\title{{\it{Ab-initio}} GMR and current-induced torques in Au/Cr multilayers}
\author{P.M. Haney$^{1}$}
\email{haney411@physics.utexas.edu}
\author{D. Waldron$^{2}$}
\email{derek.waldron@mail.mcgill.ca}
\author{R.A. Duine$^{3}$}
\email{duine@phys.uu.nl}
\author{A. S. N\'u\~nez$^{4}$}
\email{alvaro.nunez@ucv.cl}
\author{H. Guo$^{2}$}
\email{guo@physics.mcgill.ca}
\author{A.H. MacDonald$^{1}$}
\email{macd@physics.utexas.edu}
\homepage{http://www.ph.utexas.edu/~haney411/paulh.html}
\affiliation{$^{1}$Department of Physics, The University of Texas
at Austin, 1 University Station, Austin TX 78712}
\affiliation{$^{2}$Centre for the Physics of Materials and
Department of Physics, McGill University, Montreal PQ, H3A 2T8,
Canada} \affiliation{$^{3}$Institute for Theoretical Physics,
Utrecht University, Leuvenlaan 4, 3584 CE Utrecht, The
Netherlands} \affiliation{$^{4}$Instituto de F\'isica, PUCV Av.
Brasil 2950, Valpara\'iso Chile}
\date{\today}

\begin{abstract}
We report on an {\em ab-initio} study of giant magnetoresistance
(GMR) and current-induced-torques (CITs) in Cr/Au multilayers that
is based on non-equilibrium Green's functions and spin density
functional theory.   We find substantial GMR due primarily to a
spin-dependent resonance centered at the Cr/Au interface and
predict that the CITs are strong enough to switch the
antiferromagnetic order parameter at current-densities $\sim 100$
times smaller than typical ferromagnetic metal circuit switching
densities.
\end{abstract}

 \maketitle
\section{Introduction}
\label{sec:intro} Magnetic metals are often well described using
the effective mean-field description provided by the Kohn-Sham
equations of spin-density functional theory.  In this description,
the Kohn-Sham quasiparticles experience exchange-correlation
potentials with a spin dependence that is comparable in strength
to band widths and other characteristic electronic energy scales.
The spin-dependent part of the Kohn-Sham quasiparticle potential
acts like an effective magnetic field that is locally aligned with
the electron spin-density.  Because of these strong spin-dependent
potentials, the resistance of a ferromagnetic metal circuit will
change substantially when the magnetization orientation in any
part of the circuit is altered, an effect known as giant
magnetoresistance (GMR).  Conversely, transport currents can
destabalize magnetization configurations that are metastable in
the absence of a current and change collective magnetization
dynamics.  In the case of circuits containing ferromagnetic
elements the influence of transport currents on the magnetization
can be understood as following from the conservation of total spin
angular momentum; the torques that reorient quasiparticle spins as
they traverse a non-collinear magnetic circuit are accompanied by
current-induced reaction torques (CITs) that act on the magnetic
condensate. This type of phenomenon is not by any means limited to
magnetic systems. For example, there have been recent studies of
the interaction between transport and charge density waves (CDW),
which find that the CDW order parameter configuration influences
transport and conversely that transport can alter the
CDW.\cite{Mierzejewski}

GMR and CITs have been extensively studied, both experimentally
and theoretically, in ferromagnetic metal
circuits.\cite{SMT-theory,SMT-exp}  In the case of ferromagnets
current-induced torques are generally referred to as spin-transfer
torques, since they represent a transfer of conserved total spin
angular momentum between the magnetic condensate and quasiparticle
currents.  The spin-transfer argument is very general but is
perhaps a little vague in that it does not always specify what is
meant by the magnetic condensate, ({\em the magnetization}).
(Some work does appeal explicity to a {\em s-d} picture of
transition metal magnetism with {\em d}-electron local moments
exchange coupled to itinerant {\em s}-electron bands.)

Several of us have recently proposed\cite{ourafm} that CITs can
also act on the order parameter of antiferromagnetic metals.   Our
proposal follows from a microscopic picture\cite{nunezssc} of CITs
that applies to a metal with any kind of magnetic order, including
antiferromagnetic order, and suggests that CITs are a universal
phenomenon in magnetic metals. In this picture CITs arise from the
dependence on bias voltage of the relationship between the
steady-state Kohn-Sham quasiparticle density matrix and the
Kohn-Sham single-particle Hamiltonian.  This picture in effect
explicitly identifies the energy region between the chemical
potentials of source and drain, the transport window, with the
transport electrons and the energy region below this with the
magnetic condensate. Given the Kohn-Sham Hamiltonian, the
condensate contribution to the density matrix can be
constructed\cite{haney2006} (for slow condensate dynamics) by
solving the time-independent Schroedinger equation for electrons
in a system of interest including if appropriate their coupling to
electrodes in equilibrium, whereas the transport contribution to
the density matrix is constructed by solving the Schroedinger
equation with the scattering-theory boundary condition that the
electrons be incident from the source. All this is accomplished
conveniently using a non-equilibrium Greens function
theory.\cite{haney2006} The change in condensate dynamics, or the
current induced torques, follow from the difference between the
Kohn-Sham Hamiltonian constructed from the equilibrium
density-matrix and the Kohn-Sham Hamiltonian constructed from the
density-matrix in the presence of a transport voltage.  For
magnetic condensate dynamics,\cite{haney2006} what is relevant is
changes in the spin-dependent parts of the Hamiltonian and
in-particular changes in the direction of the exchange-correlation
effective magnetic fields on each site.

Several of us have recently used this approach to explore the
possibility of CITs in circuits containing antiferromagnetic
metals \cite{ourafm} by studying a simple one-band,
Hubbard-interaction, toy model. For this model, we found CITs that
drive the antiferromagnetic order parameter which are, remarkably,
proportional to film thickness, provided that inelastic
scattering\cite{inelastic} is ignored. (In ferromagnets
spin-transfer torques saturate at a finite value for large
ferromagnetic film thicknesses.) In this paper, we evaluate CITs
for a potentially realistic magnetoelectronic system using an {\it
ab-initio} non-equilibrium Green's function
formalism.\cite{reviews}  This calculation convincingly
demonstrates that GMR and CITs do occur in circuits containing
only antiferromagnetic and paramagnetic elements, and that they
can be large.  The {\em ab-initio} approach accounts for all
electronic structure details of the materials and the interfacial
bonding between materials, while the NEGF formalism enables the
calculation of finite bias properties, such as current, and
non-equilibrium spin densities.  The specific calculations we
report on in this paper were performed on a system with
antiferromagnetic (100) growth direction {\em bcc} Chromium (Cr)
leads separated by a {\em fcc} Gold (Au) spacer.  The
Au(100)/Cr(100) multilayer system we consider here appears to be
attractive as a model system for antiferromagnetic metal
spintronics.  Au/Cr multilayers and (100) growth direction epitaxy
were studied some time ago both experimentally\cite{zajac} and
theoretically,\cite{fuPRB} motivated in part by superconductivity
that can occur in disordered Au/Cr films. We find that both GMR
and CITs do occur in this ferromagnet-free magnetoelectronic
circuit, as anticipated by previous work.\cite{ourafm}  Our
calculations also identify new physics not anticipated in the
early toy-model study. The new features are associated with
spin-polarized interface resonance at the Au/Cr interface and with
the presence of more than one propagating Cr channel at the Fermi
energy for some transverse wavevectors.
The paper is organized as follows: in Sec.~\ref{sec:negf}, we
briefly review our calculation method, while in
Sec.~\ref{sec:details}, we give specific details of the system
under consideration.  We explain our results in
Sec.~\ref{sec:results} and finally, in Sec.~\ref{sec:conclusions}
we present our conclusions and suggest some possible directions
for future research directed toward exploring the potential of
spintronics in Au/Cr and related systems.

\section{Method}
\label{sec:negf} To calculate equilibrium and non-equilibrium
properties of the system, we employ the non-equilibrium Greens'
functions (NEGF) formalism within the density functional theory
framework.\cite{taylor}  The central object in this formalism is
the equal time lesser Green's function, defined as
$G_{j,j'}^{<}(t,t) \equiv i\langle c^{\dag}_{j'}(t)c_{j}(t)
\rangle$.  Here the labels $j,j'$ refer to the single-particle
basis set in which the non-equilibrium density matrix is
evaluated.  In the NEGF formalism one of the labels must specify
an atomic site, and the other labels specify included spin and the
orbitals included at each site.  These calculations normally
assume transverse periodicity, either on an atomic scale as in
this calculation, or in the context of a transverse super-cell
model for transport through a system with a finite cross section.
The single-particle basis therefore also includes a transverse
wavevector label.  The NEGF formalism implementation employed here
has been described in detail in previous work.\cite{taylor} In the
present work we assume that the time-dependence of the mean-field
Hamiltonian, which is associated with slow condensate dynamics
when present, can always  be ignored in using the formalism to
calculate a steady state value of $G_{j,j'}^{<}(t,t)$.  This
steady state Greens function specifies the non-equilibrium
density-matrix $\rho$ of the system.

 Once $\rho$ is determined for a particular magnetization orientation configuration, the GMR is
obtained by finding the spin-dependent transmission: $T_{\sigma} =
{\rm Tr}[({\rm Im}(\Sigma^r_L) G^r {\rm Im}(\Sigma^r_R)
G^a)_{\sigma}]$.  Here $G^{r(a)}$ is the retarded (advanced)
Green's function for the device, while $\Sigma^r_{L(R)}$ is the
self energy which accounts for the semi-infinite left (right)
lead.

This is the first {\em ab-initio} evaluation of either GMR or CITs
in a magnetoelectronic circuit that does not contain ferromagnetic
elements.  CITs in antiferromagnetic systems cannot be estimated
on the basis of total spin conservation considerations because the
antiferromagnet order does not carry a total spin.  The CITs
must be computed microscopically by finding the change in the
local spin-dependent exchange-correlation potential on each site
due to the presence of non-equilibrium, current carrying
electrons, and summing over all sites assuming rigid
antiferromagnetic order.  Rigidity is maintained because the
torques associated with short-distance-scale changes in relative
spin orientation are very large compared to CITs.  The CITs are
significant experimentally because they compete only with much
smaller anisotropy torques. This picture of current induced
torques, and its implementation in the NEGF formalism, is spelled
out in more detail in Ref.~\onlinecite{haney2006}.

The contribution of atom $i$ to the CIT (per current) is:
\begin{eqnarray}
\frac{{\vec {\Gamma}}_i}{I} = \frac{\mu_B}{e} \frac {\int
dk_\parallel \sum_{\alpha}({\vec \Delta}_{i,\alpha} \times {\vec
s}_{i,\alpha})} {\sum_\sigma \int dk_\parallel T_{\sigma}(E_f)}.
\label{CITlocal}
\end{eqnarray}
In the above $\alpha$ refers to an orbital, so that the sum is
over all orbitals $\alpha$ of atom $i$.  ${\vec \Delta}_{i,\alpha}
\cdot \vec{\tau}/2$ is the spin-dependent part of the
exchange-correlation potential for orbital $\alpha$ on atomic site
$i$; its magnitude is in effect the size of the spin-splitting for
this orbital and site.  Here $\vec{\tau}$ is the vector of Pauli
spin matrices. (The presence of more than one orbital on a site
and the orbital dependence of the spin-splitting fields they
produce is one of the important differences between more realistic
models of transition metal magnetism and simple generic 1-band toy
models of ferromagnetism.)  The orbital and site-dependent
non-equilibrium spin density which appears in Eq.(~\ref{CITlocal})
is given by
\begin{eqnarray}
{\vec s}_{i,\alpha}={\rm Tr}[ \rho^{(j,j)}_{tr}{\vec \tau}].
\qquad j=(i,\alpha)
\end{eqnarray}
with the trace in spin space.   Since the system we consider is
metallic, we assume that the non-equilibrium quantities in which we are
interested can be evaluated in linear
response; specifically for the non-equilibrium density matrix:
\begin{eqnarray}
\rho_{tr}=G^r{\rm Im}(\Sigma_L^r)G^a.
\end{eqnarray}
with all quantities evaluated at the Fermi energy. The torque
which acts on the staggered antiferromagnetic order parameter is
the corresponding sum of the torques on individual atoms. Since
the directions of the spin-dependent exchange-correlation fields
alternate from site to site, current induced transverse
spin-densities of the same sign give total torque contributions of
the same sign.\cite{ourafm}

\section{Cr/Au/Cr Multilayers}
\label{sec:details}  We study a circuit with semi-infinite
antiferromagetic Cr leads and a Au spacer.  Cr has a {\em bcc}
lattice structure with lattice constant 2.88 \AA, while Au is {\em
fcc} with lattice constant of 4.08 \AA.  The interface between
these materials has a fortuitous lattice matching of
two-dimensional square nets when they are grown epitaxially along
the [001] direction and the Au lattice is rotated by $45^{\circ}$
around the growth direction. In this configuration the bulk square
net lattice constants differ by less than .2\%.  The lattice
matching strains for Au on a [001] Cr substrate are therefore
quite small.

The antiferromagnetic state of bulk Cr has been studied
extensively.\cite{fawcett} The origin of antiferromagnetism in Cr
is nesting between electron jack and hole neck pieces of the
paramagnetic Fermi surface.\cite{lomer, overhauser} The nesting
vector ${\bf Q}$ defines the spin density wave (SDW) period which
is nearly commensurate with the lattice with ${Q}a/2\pi=.95$,
where $a$ is the lattice constant.  In thin film and multilayer
structures, Cr can exhibit paramagnetism as well as commensurate
and incommensurate SDW states, depending on the film thickness and
on the adjacent
materials.\cite{zabel,fullerton,fishman,pierce,fullertonPRL} There
is evidence of antiferromagnetism in Cr thin films grown on Au
substrates for coverages greater than 12 ML.\cite{O'neill} Density
functional theory, within LSDA, has been previously used to study
bulk Cr in a commensurate SDW state\cite{hafner}, and can predict
the magnitude of the exchange splitting, the magnetic moment, and
the lattice constant.  For this study, we restrict our attention
to Cr with a commensurate spin density wave structure in which the
body center spins and cube corner spins have opposite
orientations. This magnetic structure is metastable in the absence
of a bias voltage for our thin film structures in the local
spin-density approximation. All interfaces are perpendicular to
the [001] direction.

For the GMR calculations discussed below
we have used a double-zeta with polarization
basis set for both Cr and Au, and have found excellent agreement with
bulk band structures and density of states calculations.  For the
calculation of current induced torques, we have used a single-zeta
with polarization basis set, which still retains good accuracy for
bulk properties. For the calculation of the self-consistent equilibrium
density matrix, contributions from 900 $k$-points within the 2-d Brillouin zone have been summed.
For the calculation of the conductance and current induced
torques, 25600 $k$-points have been summed.

As a matter of convention, we define parallel (P) and
anti-parallel (AP) alignment below in terms of the alignment of
the Cr spins in the two layers on opposite sides of the Au spacer.
The inset of Fig. 1 shows the geometry and spin structure of the
systems considered.

\begin{figure}
\vskip 0.25 cm
\includegraphics[width=3.5in,angle=0]{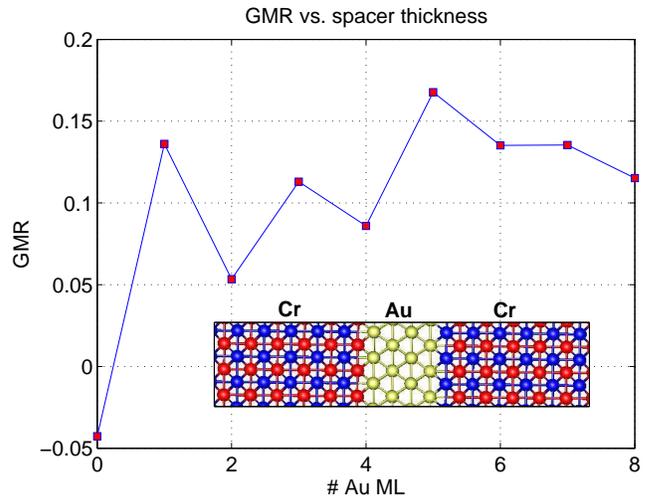}
\vskip 0.25 cm \caption{GMR as a function of spacer thickness.
There is a sizable GMR for all spacer thicknesses.  The inset
shows the geometry for the four layer spacer configuration - up
and down spins are colored red and blue, respectively.  The
configuration above illustrates an anti-parallel configuration. }
\label{bands}
\end{figure}

\section{Results}
\label{sec:results}  Fig 1. illustrates the dependence of the GMR
ratio (defined as $(I_P-I_{AP})/I_{AP}$) on spacer thickness. In
the limit of no spacer, the GMR ratio is negative, implying larger
conductances for an antiparallel configuration. This property is
anticipated since the AP case at zero spacer thickness corresponds
simply to ballistic conduction through bulk antiferromagnetic Cr
with unit transmission coefficient for all channels, whereas the
parallel configuration implies a kink in the Cr antiferromagnetic
order parameter configuration which reduces the transmission. For
all nonzero spacers we have studied, we find that the GMR ratio is
positive. The nonzero GMR for antiferromagnetic systems is perhaps
surprising at first sight; for example a simple Julliere type
two-channel conductor model, in which MR is due to spin-dependent
conductance in the bulk, would predict that the GMR ratio is zero
for antiferromagnetic systems. For antiferromagnets GMR is, in
this sense, purely an interface effect; for ferromagnets GMR is
only partly (but often mainly) an interface effect.
\begin{figure}
\vskip 0.25 cm
\includegraphics[width=3.5in,angle=0]{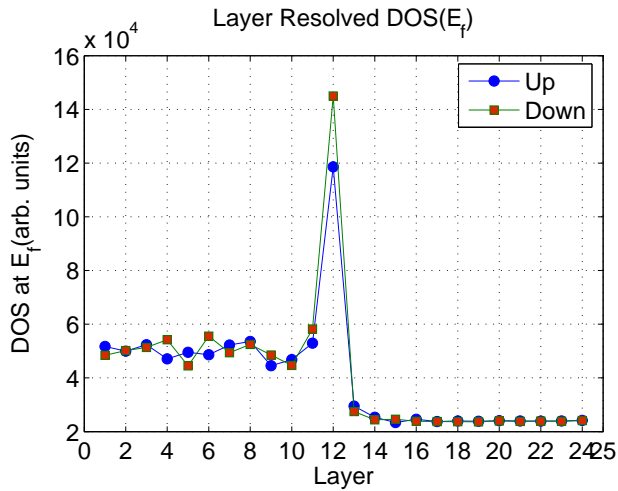}
\vskip 0.25 cm \caption{Layer and spin resolved density of states
at the Fermi energy for Cr-Au single interface system. Layers 1-12
are Cr, and 13-24 are Au.  The pronounced enhancement of the
density-of-states in layer 12 reflects the interface resonance.
The spin-polarization of the interface resonance is consistent
with the fact that $T_\downarrow>T_\uparrow$ in the transport
calculation.} \label{bands}
\end{figure}

In the case of the previous toy model antiferromagnetic
calculations\cite{ourafm}, GMR was due to phase coherent multiple
scattering between two antiferromagnets.  These effects are
partially mitigated\cite{inelastic} at elevated temperatures by
inelastic scattering which breaks phase coherence.  The present
{\em ab-initio} calculations reveal a new contribution to
antiferromagnetic GMR, explained below, which does not rely on
phase coherence. The property that realistic antiferromagnets have
GMR effects that are not dependent on phase coherence is
encouraging from the point of view of potential applications,
since it suggests larger robustness at elevated temperatures.

In order to identify the dominant GMR mechanism of Au/Cr
multilayer systems, we have performed a separate NEGF calculation
for a single interface between semi-infinite bulk Cr, and
semi-infinite bulk Au. The result is that there is a spin
dependent conductance with magnitude
$(I_\uparrow-I_\downarrow)/(I_\uparrow+I_\downarrow) = -2.10\%$.
The current is spin-polarized in the direction opposite to the top
layer of the antiferromagnet. For Cr/Au/Cr multilayers, this
spin-filtering implies that the conductance is maximum when the
facing layers of the antiferromagnet have the same
spin-orientation, {\em i.e.} the P configuration has a higher
conductance, even without any non-local coherence effects. This
effect is absent\cite{ourafm} in the single-band models that we
studied earlier.

\begin{figure}
\vskip 0.25 cm
\includegraphics[width=3.5in,angle=0]{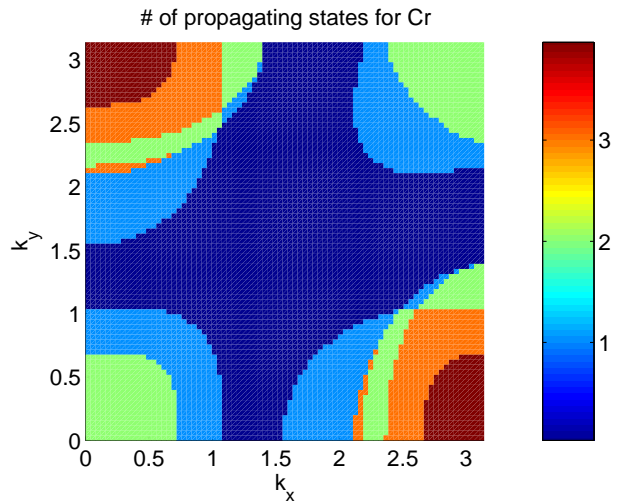}
\vskip 0.25 cm \caption{Number of propagating states in the [001]
direction at the Fermi energy for bulk antiferromagnetic Cr.}
\label{bands}
\end{figure}

To explore the origin of the spin-dependent interface resistance
in greater detail, we have examined the layer and $k_{\|}$
resolved local density of states of the single-interface
calculation (here $k_{\|}$ refers to the transverse momentum
label).  Fig. 2 shows the layer-resolved results. (In the
calculation, the first 12 layers on each side of the interface are
allowed to differ from the bulk.  In Fig. 2 layers 1-12 are the Cr
layers in the scattering region near the interface while layers
13-24 are the Au layers in the scattering region.) We see that
there is a pronounced interface resonace on the last Cr layer;
this is a consequence of the difference between the Fermi surface
topologies of Cr and Au. Moreover, this state is spin polarized,
with direction opposite to that of the bulk local moment. Fig. 3
shows the number of propagating channels in the Brillouin zone for
Cr, which demonstrates that the Fermi surface of Cr differs
strongly from the nearly spherical Fermi surface of Au.
In particular, there are large regions in the Brillouin zone of Cr
in which there is no propagating state, whereas Au has propogating
states across all of the central region of the transverse
Brillouin zone.   Fig. 4 shows the transverse wavevector-resolved
Fermi-level local density of states for layers 8, 10, 12 (the last
Cr layer), and 16. Layer 8 is typical of bulk Cr, while layer 16
is typical of bulk Au. Layer 12, however, shows features of both
materials; in particular, populations of states within the region
of the Cr Brillouin zone with no propagating modes are responsible
for the localized interface resonance.  Fig. 5 shows the total
local-density of states as a function of energy for layers near
the interface.  The rapid relaxation toward bulk values away from
the the interface is apparent.  The interface layer has a highly
distorted density of states function, a high density of states at
the Fermi level, and a higher moment density which is responsible
for a net ferromagnetic moment\cite{fuPRB} contribution from the
interface region.  Apparently interruption of antiferromagnetic
order both narrows the majority-spin bands and lowers the energy
of minority spins in this interface layer.  Hopping of down-spins
from the sub-interface layer on which they are the majority to the
spacer layer is enhanced by the minority spin interface resonance.

The enhanced moment density in the interface layer is accompanied
by more attractive spin-up potentials on this layer and
spin-dependent bonding across the Cr/Au interface. The effective
hopping matrix elements across the interface have a spin dependent
contribution that is about $1\%$ of their total values.  In order
to determine whether it is spin-dependent hopping or resonances
related to spin-dependent site energies we have symmetrized
hopping of the interface to remove its spin dependence, and
re-calculated the conductance.  We find the same value for the
polarization, indicating that it is the interface resonance that
is largely responsible for the polarization.  The fact that the
GMR is due to interface resonances, rather than to phase-coherent
multiple-scattering across the spacer layer, suggests that the
effect will be robust at elevated temperatures.

As mentioned earlier, the antiferromagnet/normal interface
resistance is not spin-dependent in the toy model systems
previously studied.\cite{ourafm}  The key property of the toy
model which leads to this spin-independent interface resistance is
that each antiferromagnetic unit cell is invariant under a
combination of space and spin inversion.  The {\em ab-initio}
mean-field Hamiltonian does not have this property.  A
spin-dependent resistance will occur in the toy model when either
the hopping amplitude from spacer to the top antiferromagnetic
layer is made spin-dependent or the exchange-splitting in the top
layer is shifted from its bulk value.

We have evaluated the current induced torques for a system with a
four Au monolayer spacer.  The angle between the staggered moments
of the Cr leads was initialized to $90^{\circ}$: the staggered
magnetization is along the $\hat z$ direction in the Cr layers to
the left of the spacer and along the $\hat x$ direction in the Cr
layers to the right. A self-consistent non-collinear solution to
the Kohn-Sham equations was obtained with this configuration.  The
resulting layer resolved torques, evaluated as described in Sec.
~\ref{sec:negf}, are plotted in Fig. 6. We find strong torques
peaked in the first few Cr layers, in contrast to the toy model
case in which the torques were constant in magnitude and
alternated in direction from layer to layer.
\begin{figure}
\vskip 0.25 cm
\includegraphics[width=3.5in,angle=0]{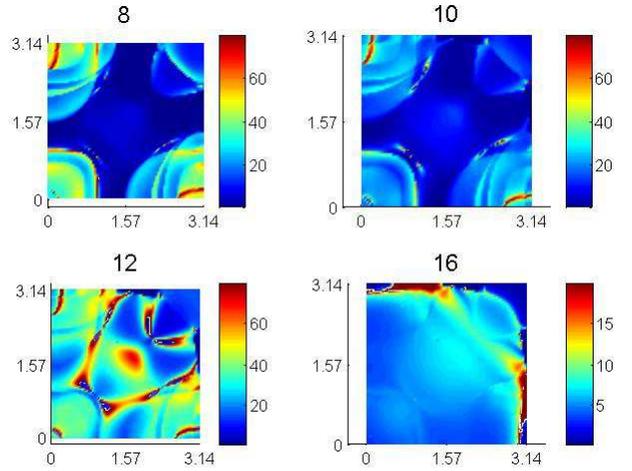}
\vskip 0.25 cm \caption{Transverse-momentum resolved Fermi-energy
density-of-states 8, 10, 12, and 16 of the single interface Cr-Au
system.  Layer 12 is the Cr layer closest to the interface. Layer
8 shows bulk Cr characteristics (compare to Fig. 3), while layer
16 shows bulk Au characteristics. } \label{bands}
\end{figure}

\begin{figure*}[h!]
\vskip 0.2 cm
\includegraphics[width=7in]{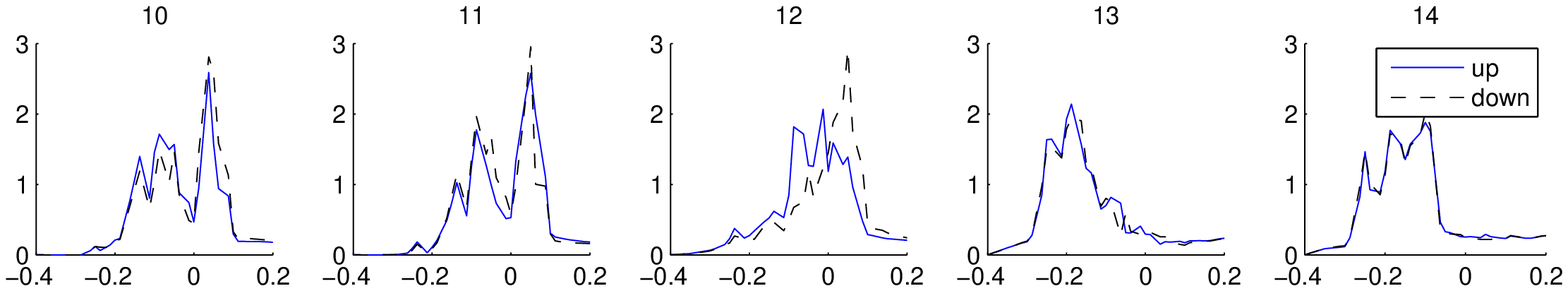}
\vskip 0.2 cm \caption{Density-of-states for layers 10-14 of the
single interface Cr-Au system.  The DOS relaxes to its bulk shape
a couple of layers away from the interface.  Layer 12 is the
interface Cr layer, end the Fermi energy is 0.} \label{bands}
\end{figure*}

To shed light on the origin of the new antiferromagnet
current-induced-torque physics revealed by these {\em ab-initio}
calculations we focus on differences between the toy-model case,
in which analytic calculations are possible, and the realistic
Au/Cr case.  Because the equilibrium torques that impose the
antiferromagnetic order will always be much stronger than the
current-induced torques, we are mainly interested in the sum of
the layer resolved torques which drives the antiferromagnetic
order parameter and therefore competes only with anisotropy
torques. The perfectly staggered torque obtained in the toy model
case arises from an out-of-plane current-induced spin density that
is spatially constant within each antiferromagnet. (Here in-plane
refers to the plane spanned by the orientations of the two
antiferromagnetic layers and out-of-plane refers to the
perpendicular direction, the $\hat y$-direction in our case.)  The
constant out-of-plane current-induced spin density in the toy
model can be partially explained by the fact that Bloch wave
vectors of up and down spin states are not spin-split in
antiferromagnets. It follows that a linear combination of
transmitted up and down spins have a transverse spin density that
is position-independent. (In contrast to the ferromagnetic case,
in which the transverse spin density of a particular channel shows
spatial oscillations with a period given by
$(k_\uparrow-k_\downarrow)^{-1}$).\cite{stiles} To see where this
physics breaks down in our calculation, consider Fig. 3, which
shows that a particular transverse channel has 1-4 possible values
of $k_z$. For those channels with a single $k_z$ value at the
Fermi energy, we find that the contribution to the transverse spin
density is spatially constant.  Evidently the toy model does a
good job of describing this type of transverse channel, suggesting
that our earlier conclusion that there is a bulk contribution to
the staggered spin-torque in an antiferromagnet does have general
validity.  The present calculations emphasize, however, that there
is also an interface contribution coming dominantly from channels
with more than one $k_z$ value at the Fermi energy.  In this case
the transmitted wave function is a linear combination of states
with different $k_z$values.  These states interfere with each
other to produce an oscillating transverse spin density.  Summing
over many channels with different oscillation periods of the
transverse spin density leads to a rapid decay of the transverse
spin density, exactly as in the case for
ferromagnets.\cite{stiles}  A material with a simpler, single
valued Fermi surface would not have this interface contribution to
the total staggered torque.  The complex Fermi surface is
necessary to stabilize the AFM order via nesting in the first
place, however. In our case, the interface torque dominates the
total staggered torque. It is nevertheless substantial, totalling
.049 $\mu_B/e$.

To estimate the critical current for switching the
antiferromagnetic order parameter, we take the anisotropy of bulk
Cr with spins pointing along the ${\bf n}$ direction as: $E({\bf
n})=K_1({\bf \hat z \cdot n})^2+4K_2({\bf \hat x \cdot n})^2({\bf
\hat y \cdot n})^2$, where\cite{Fenton} $K_1=10^3$ J m$^{-3}$, and
$K_2=10$ J m$^{-3}$, and take the magnetic damping parameter to be
$\alpha=0.1$.\cite{Fenton}  Here $K_1$ is positive for
$T>T_{sf}=123.5 K$, and ${\bf Q}$, the spin density wave vector,
is taken to be in the z-direction.  Near the fixed point, ${\bf
n}={\bf \hat x}$, the damping torque per area is then
$\Gamma=\alpha\gamma(K_1+4K_2)t$, where $t$ is the thickness of
the layer.  (Note that antiferromagnets possess no demagnetizing
field, so that the anisotropy does not include shape anisotropoy
and is due only to magnetocrystalline anisotropy.) The current
required to have the current-induced-torque overcome damping is
therefore $.049(\mu_b/e)\Gamma \approx 6.3 \cdot 10^{18}t
(A/m^3)$.  (In this we assume that the transfer torque efficiency
calculated at $\theta=90^{\circ}$ is the same as that for small or
large angle.)
Typical values for critical current densities in ferromagnets are
up to 100 times larger, primarily because of the large
demagnetizing fields present in ferromagnets. We therefore expect
that it will be possible to achieve current induced switching in a
circuit containing only antiferromagnetic elements.

\begin{figure*}[h!]
\vskip 0.2 cm
\includegraphics[width=7in]{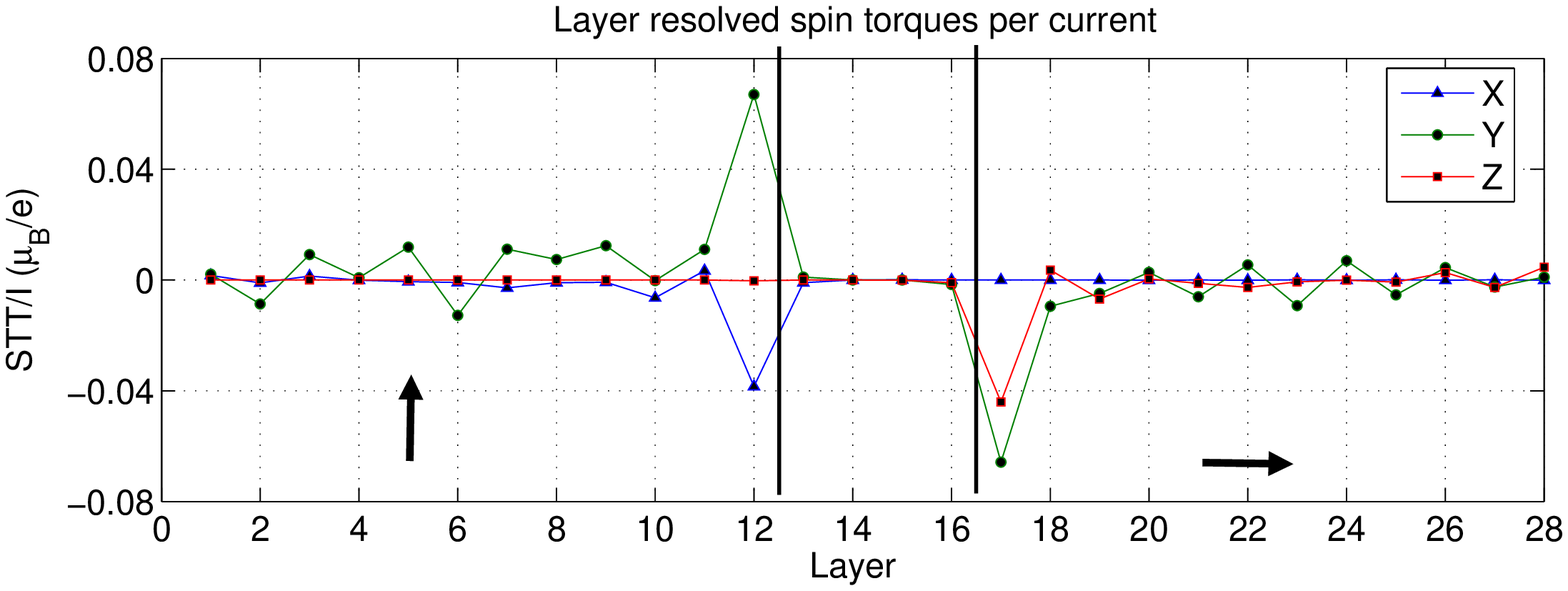}
\vskip 0.2 cm \caption{Layer resolved current induced torque per
current.  The (X,Y,Z) directions in the legend refer to the
directions of the induced torques. The torque is localized in the first
layer of the Cr. } \label{bands}
\end{figure*}

\section{Summary and Conclusions}
\label{sec:conclusions} In summary, we have performed a
first-principles calculation of the non-equilibrium properties of
Au/Cr multilayers.  We have found that a system composed of Cr
leads with an Au spacer possesses a robust GMR effect, which is
due primarily to a spin polarized interface state at the Au/Cr
interface, and a strong current induced torque which corresponds
to a switching critical current more than an order of magnitude
smaller than that of ferromagnets. These {\em ab-initio}
calculations demonstrate that current-induced-torques in
antiferromagnets have a bulk contribution from channels with only
one band at the Fermi energy and an interface contribution, which
is dominant in our case, from channels with more than one band at
the Fermi energy. Our calculations demonstrate that all the basic
effects of metal spintronics occur in circuits with only
paramagnetic and antiferromagnetic elements.

The robust antiferromagnetic spintronics effects studied here
occur in circuits in which current flows perpendicular to planes
containing perfectly uncompensated spins.  We believe that
substantial effects will also occur in realistic experimental
systems in which all layers are partially compensated. Because the
order in an antiferromagnet is staggered, structural disorder
(interface roughness for example) will induce magnetic
order.\cite{ravli} Although we have not studied these effects in
any depth to date, it seems likely that observable GMR and
current-induced order-parameter and resistance changes will
survive as long as the length scales associated with spatial
disorder are larger than the spacer layer thickness.  For Cr on Fe
whiskers the lateral correlation length is the terrace width,
which can be easily larger than typical spacer widths if the
interface is flat.\cite{pierce,fullertonPRL,FeCr}

It may be that initial studies of spin-torques in
antiferromagnetic systems will be easier to analyze in systems in
which the antiferromagnets are exchange coupled to ferromangets,
whose orientation can be manipulated by external magnetic fields.
There is, for example, already evidence\cite{wei} that
current-induced torques act on the antiferromagnetic layer in
spin-valve structures.
Systems containing Cr antiferromagnetic films that are exchange
coupled to Fe whiskers might also be attractive to study these
kinds of current-induced torques since it is already known that
weakly compensated Cr layers can be obtained relatively
easily.\cite{FeCr,pierce,fullertonPRL}  It may also be
advantageous to consider Mn doped Cr, as it is known to have a
higher N\'eel temperature than pure Cr, and forms a commensurate
spin density wave.\cite{fawcett2}  Although the materials
challenges presented by antiferromagnetic metal spintronics are
even stronger than those presented by ferromagnetic spintronics,
we believe that the subject will prove interesting from both basic
physics and potential application points of view.

\acknowledgments  The authors acknowledge helpful interactions
with Jack Bass, Sam Bader, Joseph Heller, Olle Heinonen, Chris
Palmstrom, Stuart Parkin, Maxim Tsoi and Z.Q. Qiu. This work was
supported by the National Science Foundation under grant
DMR-0606489, by a grant from Seagate Corporation, and by the Welch
Foundation.  ASN was partially funded by Proyecto MECESUP FSM0204.
Computational support was provided by the Texas Advanced Computing Center.

\end{document}